\documentclass[%
 reprint,  amsmath,amssymb,
 aps,
]{revtex4-2}

\usepackage{dcolumn}
\usepackage{bm}
\usepackage{graphicx}
\usepackage{subfigure}
\usepackage{bm}
\usepackage{latexsym,amsthm,amssymb}
\usepackage{booktabs,threeparttable}
\usepackage{makeidx}
\usepackage{amsmath}
\usepackage{amsfonts}
\usepackage{amssymb}
\usepackage{setspace}
\usepackage{slashed}
\usepackage{diagbox}
\usepackage{array}
\usepackage{mathtools}

\begin{document}

\preprint{APS/123-QED}

\title{ The Quantum Channels of the Compton Scattering of an Entangled Pair of Photons }

\author{Chih-Yu Chen}
  \email{ayufae@gmail.com}
\affiliation{Quantum Information Center, Chung Yuan Christian University, Taoyuan, Taiwan}

\author{C. D. Hu}
  \email{cdhu@phys.ntu.edu.tw}
\affiliation{Department of Physics, National Taiwan University, Taipei, Taiwan }

\author{Yeu Chung Lin}  
\email{albertyclin@gmail.com}
\affiliation{Department of Physics, National Taiwan University, Taipei, Taiwan}

\begin{abstract}
The Compton scattering of an entangled pair of photons is studied. Using the resulting reduced density matrix, we demonstrate how to derive the Kraus operators for the interaction from the underlying fundamental theory, quantum electrodynamics. The normalization of the final density matrix of the entangled photons after the interaction is crucial to obtaining the correct probabilities and correlations. The reduced density matrix of the photon, which does not participate in the interaction directly, remains unchanged. It indicates that quantum effects in quantum sensing using an entangled photon pair can be observed only through the correlations or superposition of the entire pair. We relate our results to the Stokes parameters for determining polarizations and investigating the entangled states of photons from electron-positron annihilation in astrophysical systems.
\end{abstract}
\maketitle 

\section{introduction}
\indent
Quantum entanglement is a state in which each particle in the system cannot be described separately. This concept is essential for illustrating the relationship between the foundations of quantum mechanics and the hidden-variable theory proposed by Einstein, Podolsky, and Rosen (EPR) \cite{EPR}. Furthermore, experimental tests of the EPR theory often involve entangled states. Photon states constitute a particularly important class \cite{Degen}. These experiments must account for photon-matter interactions, and a commonly used one is Compton scattering. Therefore, it is not surprising that one of the earliest studies of experiments concerning EPR theory was that of a pair of distantly correlated (entangled) photons, each photon undergoes Compton scattering \cite{BA}. \\
\indent
Experimental studies of this topic have a long history  \cite{Kasday}. Recently, many theoretical and experimental works \cite{Caradonna} have investigated the implications of entangled states with Compton scattering. Notably, Parashari et al. \cite{Parashari} studied the angular correlation of Compton-scattered annihilation photons and hidden variables. Tkachev et al. \cite{Tkachev} investigated polarization correlations in entangled photons, for entangled-photon pairs in which one photon had been Compton-scattered. Therefore, we developed quantum operations of Compton scattering on the density matrix.\\
\indent
Three equivalent schemes can be used to treat quantum operations between a principal system and its environment, in the current case, an entangled pair of photons and an electron: physically motivated axioms, the operator-sum representation, and a system coupled to the environment \cite{Nielsen}. \\
\indent
The first approach constructs a set of physically motivated axioms that we expect a dynamical map in quantum mechanics to satisfy. The general approach needs more insight into how to calculate the actual dynamics. The operator-sum representation, or the quantum channel \cite{Kraus} approach, is mathematically simple in applications. The third approach involves exploring the detailed dynamics between the system and its environment. One of its major applications is to use a set of operators acting on the density matrix, which represents the system's state, to describe the interaction between the system and the quantum channel's environment. Establishing a set of operators corresponding to a specific interaction by physical intuition is not particularly challenging if the interaction is simple. However, deriving a Kraus operator set from the fundamental theory of the corresponding interaction requires deliberate effort.\\
\indent
This study uses the fundamental theory of quantum electrodynamics (QED) to investigate Compton scattering of an entangled photon pair. Our focus, particularly on the entanglement, is on the dynamics of the photons’ polarizations. The entangled photons are treated as the principal system, and the electron is treated as the environment. The photon that interacts directly with the electron is dubbed the signal photon; the other is dubbed the idler photon, as it is not directly involved in the interaction. \\
\indent
Unlike the conventional approach to QED Feynman amplitude calculations, we use the density matrix, which represents the polarizations of the entangled photons, to calculate the interaction process and retain only the polarization density matrix after Compton scattering. QED is a perturbative theory. The initial state receives the zeroth order (the noninteracting part) and the first order (the interacting part) of the quantum correction, becoming the final state. To maintain the unitarity of states, the final states must be normalized.\\
\indent
In density matrix calculations, the unitarity of states is typically enforced by requiring the trace of the density matrix to be one; for instance, see Eq. (10) in Ref. \cite{Araujo}. However, a generic density matrix for entangled photon pairs includes matrix elements corresponding to different states, which may experience varying interaction strengths and require different normalizations after the interaction. In this regard, we develop a normalization scheme for the density matrix of an entangled photon pair. The established procedure is crucial to obtaining the correct probability and correlation elements of the density matrix.\\
\indent
The quantum channels for Compton scattering between the principal system and its environment are derived by comparing the initial and properly normalized final polarization density matrices of the entangled photons. One of the main results of this work is the detailed calculation of the process deduction from a fundamental theory to a set of Kraus operators corresponding to a specific interaction.\\
\indent
Using the derived density matrix and the quantum channel set, we explore the physics of the effective action of Compton scattering on the polarizations of entangled photons. It is found that Compton scattering behaves as a depolarizing mode for polarizations when considering only the horizontal polarizations. The entanglement entropy decreases after Compton scattering primarily due to depolarization. The reduced density matrix of the signal and idler photons is derived to characterize their mutual information. The mutual information of the system is also reduced due to the correlation reduction in the Compton scattering process.\\
\indent
It has been shown that photons can travel an astronomical distance without losing their coherence \cite{Berera}. We investigate the Compton scattering of a pair of entangled photons created by electron-positron annihilation in astrophysical systems. Furthermore, we present our results in terms of Stokes parameters for optical study. \\
\indent
The article is organized as follows. Section II investigates the single-photon polarization degree of freedom in Compton scattering in a density matrix form. We adopt a normalization scheme in which each element in the final density matrix is normalized by its corresponding normalization constant. This scheme differs from the ordinary normalization scheme, which divides the final density matrix by an overall normalization constant derived from the sum of its diagonal elements. In Section III, we study Compton scattering of an entangled photon pair and calculate the final density matrix of the photons. We continue to adopt the aforementioned normalization scheme and prove that it yields the correct physical results. Section IV derives the Kraus operators for single-photon and entangled-photon states corresponding to the Compton scattering. We compute the reduced density matrix for the signal and idler photons. Using them, we derive the entropy and mutual information. We explore the physical results from these calculations. In Section V, the relation between our reduced density matrix and Stokes parameters is established, and the entanglement of photons from astrophysical sources is studied. Finally, we conclude and summarize our findings in Section VI. \\
\section{photon polarization}
\begin{figure}[t]
\centering
\includegraphics[scale=0.26]{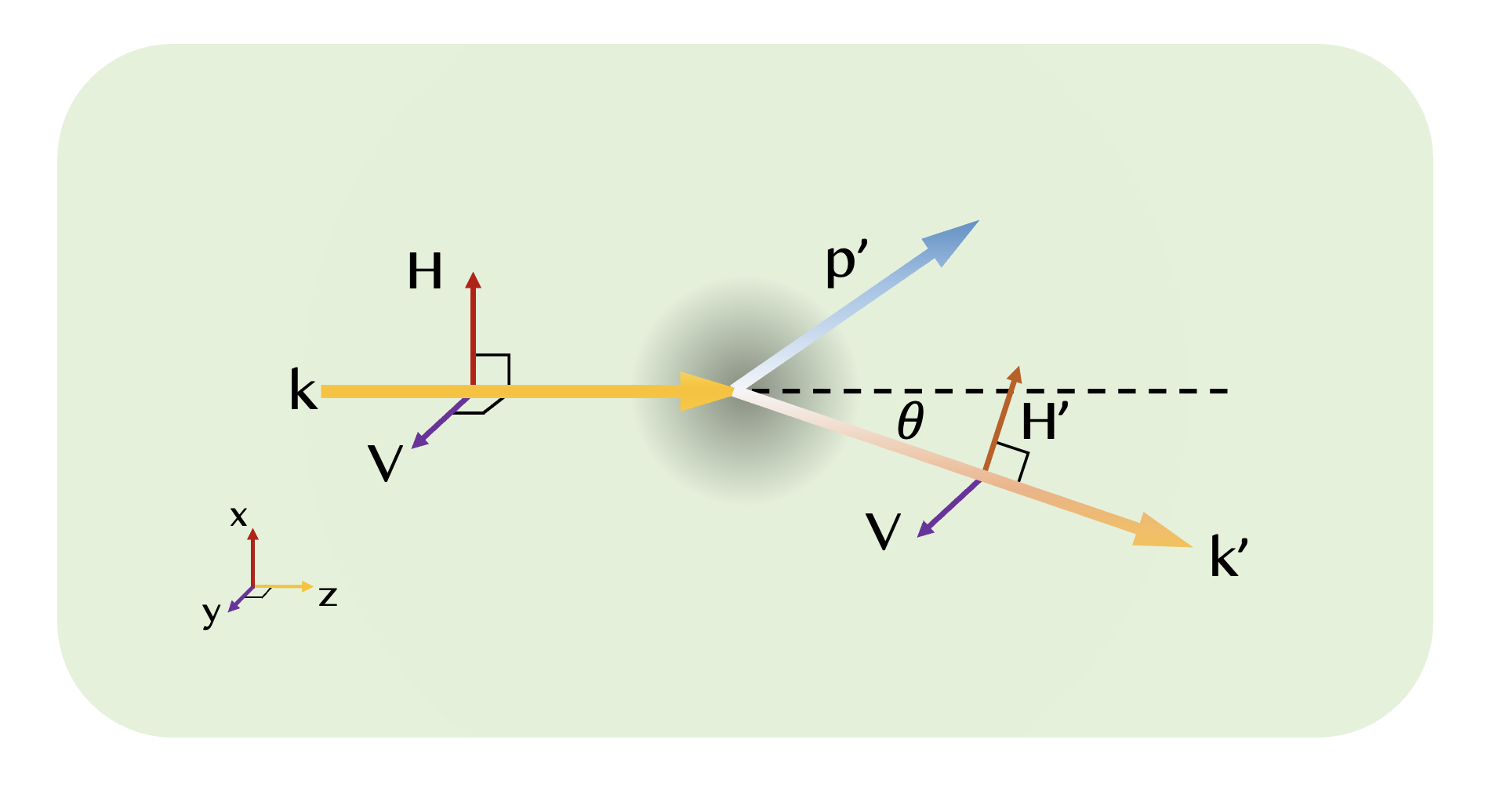}
\caption{ The illustration of the Compton scattering. An incoming photon with momentum $\mathbf{k}$ becomes $\mathbf{k'}$ after the scattering. The scattering happens on the $xz$-plane, which is denoted by index $H$ or $H'$. The directions of $V$ ($y$-axis) are vertical to the scattering plane. The electron momentum is denoted by $\mathbf{p'}$. }
\label{figCompton}
\end{figure}

We first study single-photon Compton scattering to illustrate the general calculation procedure using the density matrix. The famous Klein-Nishina formula for Compton scattering \cite{Sakurai} is
\begin{equation}\label{KNformula} 
\frac{d\sigma}{d\Omega}=\frac{r^2_0}{4}\left(\frac{\omega'}{\omega} \right)^2 \left( \frac{\omega'}{\omega} + \frac{\omega}{\omega'}-2+4(\hat{\varepsilon}\cdot \hat{\varepsilon}' )^2 \right),
\end{equation}
where $r_0=\frac{e^2}{4\pi \epsilon_0 mc^2}$ is the classical electron radius, $\omega$ $(\omega')$ and $\hat{\varepsilon}$ $( \hat{\varepsilon}')$ are the frequencies and
polarizations of the incident (scattered) photons, respectively. Let $\theta$ be the angle between the incident and scattered photon momenta. We have
\begin{equation} 
\omega'=\frac{\omega}{1+\omega(1-\cos\theta)/m}.
\end{equation} 
\indent
The scattering is depicted in FIG. \ref{figCompton}. An incident (signal) photon with momentum $\mathbf{k}$  collides with an electron at rest in the lab frame. After the collision, the scattered photon has momentum $\mathbf{k}'$, and the electron has momentum $\mathbf{p}'$. For convenience, we set the photon's incident momentum to be in the $z$-direction, and the scattering plane to be the $xz$-plane. Then the $\hat{\varepsilon}$ is in the $y$-direction for $V$ and in the $x$-direction for $H$ whereas $\hat{\varepsilon}_{H'}$ is in the $xz$-plane and has a $z$-component. It is written as $H'$ to distinguish it from the polarization of the incident photon $H$. The distinction is necessary because the wave vectors of the incident and scattered photons have different directions. Calculating the cross section and density matrix is straightforward with this basis. Despite the fact that $H$-$V$ and $H'$-$V$ planes are two-dimensional in nature, we choose a three-dimensional representation to keep track of the change of initial and scattered momenta, so that the initial and final polarization can be incorporated in the same space. Interested readers can find the details in Appendix A \cite{CHLSM}.\\
\indent
This choice of coordinates is convenient for calculation, as it is explained in the third line below Eq. (B1) in Appendix B \cite{{CHLSM}}. It is different from setting a laboratory coordinate frame prior to scattering experiments. In that case, the direction of the incident photon is also set to be the $z$-axis. The scattered photons emerge at various azimuthal angles $\phi$. Fortunately, the differential cross section is independent of $\phi$ as shown in Eq. (\ref{KNformula}). Hence, our results can be applied to any value of $\phi$ or the sum of the experimental data of all values of $\phi$.\\
\indent
The initial state of the photon-electron is written as $\left | i \right>^{sgl} = \left | 0, a \right> \left | k, \hat{\varepsilon} \right> $ where $0$ in the first bracket on the right-hand side indicates the electron is at rest, $a$ denotes the electron spin and $\hat{\varepsilon}$ denotes the polarization of the incident photon. The final state will be
\begin{equation}\label{finalstate} 
\begin{aligned}
\left | f_{\hat{\varepsilon}} \right>&^{sgl} =\\
& \frac{1}{\sqrt{ N_{\hat{\varepsilon}}}}  \big[  \left | i \right>^{sgl}  + i (2\pi)^4 \sum_{a'} \int_{p', k'}  \delta^{4}(k-p'-k')\times\\
& M(0, a, k, \hat{\varepsilon} ; p', a', k', \hat{\varepsilon}')  \left | p', a' \right>  \left | k', \hat{\varepsilon}' \right>  \big],
\end{aligned}
\end{equation}
where $p’, a’ $ are the momentum and spin of the final state of the electron, $k’, \hat{\varepsilon}'$ are the momentum and polarization of the scattered photon. The matrix $M(0, a, k, \hat{\varepsilon} ; p', a', k', \hat{\varepsilon}') $ of the lowest order Compton scattering shows the initial states and the final states of the electron and photon separated by a semicolon. $\int_{p'}=\int d^3 p'/[(2\pi)^3 2E_{p'} ]$ and $\int_{k'}=\int d^3 k'/[(2\pi)^3 2\omega']$ sum over all the possible states with $E_{p'}$ and $\omega'$ being the energies of the scattered electron and photon. In Eq.(\ref{finalstate}), the factor $ 1/\sqrt{ N_{\hat{\varepsilon}}} $ is the normalization constant, which normalizes the final polarization with the initial polarization state $\varepsilon$. It is important to normalize the wave function as required by quantum mechanics. As will be shown later, omitting this step can yield erroneous results. \\
\indent
We now choose the initial state to be the superposition state of the two basis states of the polarization, so that we can probe all the quantum channels in the Compton scattering
\begin{equation}\label{initialSinglePhoton} 
\left | i \right>^{sup}  = \frac{1}{\sqrt{2}} \left( \left | 0,a \right> \left | k, \hat{\varepsilon}_H \right>  +\left | 0,a \right> \left | k, \hat{\varepsilon}_V \right>\right).
\end{equation}  
The final state density matrix is
\begin{widetext}
\begin{equation}\label{finalDM} 
\begin{aligned}
\rho^{sup}_f&= \frac{1}{2} \sum_{\hat{\varepsilon}_m, \hat{\varepsilon}_n=\hat{\varepsilon}_H, \hat{\varepsilon}_V}\left | f_{\hat{\varepsilon}_m}\right>^{sup}\left < f_{\hat{\varepsilon}_n} \right | ^{sup}  \\
&= \frac{1}{2} \sum_{\hat{\varepsilon}_m, \hat{\varepsilon}_n=\hat{\varepsilon}_H, \hat{\varepsilon}_V} \frac{1}{  \sqrt{ N_{\hat{\varepsilon}_m} N_{\hat{\varepsilon}_n}} }  \Big[ \left| 0,a \right> \left| k, \hat{\varepsilon}_m \right>  \left< k, \hat{\varepsilon}_n \right | \left< 0, a \right | \\
&\quad+ i (2\pi)^4 \sum_{a'} \int_{p', k'} \delta^{4}(k-p'-k') M(0, a, k, \hat{\varepsilon} ; p', a', k', \hat{\varepsilon}')  \left | p', a' \right>  \left | k', \hat{\varepsilon}' \right> \left< i \right |^{sup} + h.c.  \\
&\quad +  (2\pi)^8  \sum_{a', a''} \int_{p', k'} \delta^{4}(k-p'-k') \int_{p'', k''} \delta^{4}(k-p''-k'') \times \\
& \qquad M(0, a, k, \hat{\varepsilon} ; p', a', k', \hat{\varepsilon}')  M^{*}(0, a, k, \hat{\varepsilon} ; p'', a'', k'', \hat{\varepsilon}'')  \left | p', a' \right>  \left | k', \hat{\varepsilon}' \right> \left<  k'', \hat{\varepsilon}''  \right | \left<  p'', a'' \right | \Big].
\end{aligned}
\end{equation}
\end{widetext}
The electron states in the final state of the density matrix can be traced out with the following formula:
\begin{equation}\label{TrRhof1}  
Tr_{e}[\rho^{sup} _f]=\sum_b \int \frac{d^3 p} {(2\pi)^3}  \frac{1} {2E_p} \left< p, b \right |  \rho^{sup}_f \left |  p, b\right>,
\end{equation}
where $p$ and $b$ are the four-momentum and spin of the electron and $E_{p}^2 = p^2 + m^2$. After this step, the electron's degrees of freedom, or the environmental effects, are integrated out, and the system is left with only the dynamics of the principal part, in which the environment's averaged influences are incorporated. We present the resulting final density matrix with rows and columns in order of $\left| H \right>$, $\left| V \right>$, $\left| Z \right>$, where $\left| Z\right>$ denotes the polarization along the $z$-axis direction. The resulting reduced density matrix of the photon is
\begin{equation}\label{DMSinglePhoton}  
\begin{aligned}
Tr_e&[\rho^{sup} _f]=\\
& \frac{1}{2}\left(
\begin{array}{ccc}
\frac{1+3\mathcal{T}\sigma_t /10\mathcal{V}}{ N_{H}} & \frac{1+ \mathcal{T}\sigma_t /2 \mathcal{V}}{\sqrt{N_V N_{H}}} & 0 \\
 \frac{1+ \mathcal{T}\sigma_t /2\mathcal{V}}{\sqrt{N_V N_{H}}}  &  \frac{1+ 3\mathcal{T}\sigma_t /2\mathcal{V}}{N_V }  & 0 \\
0  & 0 &  \frac{ \mathcal{T}\sigma_t/5\mathcal{V} }{ N_{H}} \\
\end{array}
\right),
\end{aligned}
\end{equation}
where the total cross section $\sigma_t  = 8\pi \alpha^2/3m^2 $ with $\alpha$ being the fine structure constant, $N_V = 1 + 3\mathcal{T}\sigma_t/2 \mathcal{V}$, and $N_{H}= 1 + \mathcal{T}\sigma_t/2\mathcal{V}$. $\mathcal{T}$ and $\mathcal{V}$ are the interaction time and volume. Appendix A\cite{{CHLSM}} contains the calculation details. \\
\indent
Note that the derivation of the final polarization density matrix in Eq. (\ref{DMSinglePhoton}) performs a solid angle integration and hence the final result contains no angular dependence. In other words, it is an angle-averaged probability and correlation of polarizations in the frame spanned by $\mathbf{k}$ and $\mathbf{k'}$. The final density matrix preserves only the polarization information of the initial photon given in Eq. (\ref{KNformula}) after the Compton scattering. The final polarization density matrix for the entangled-photon case also contains no angular dependence. Also, note that different matrix elements are normalized differently. If we were to take the approach of Ref. \cite{Araujo} to calculate the reduced density matrix, all the matrix elements would have the same normalization. This obviously leads to wrong eigenvalues; see Appendix C for a detailed proof. 
\section{Compton scattering with an entangled photon pair}
We now consider the entangled states of two photons. One is the signal photon, which scatters with an electron at rest. The other is the idler photon, which is not scattered and does not directly interact with the electron. For the signal photon, we denote the state with polarization vertical to the scattering plane as $\left| k, V\right>$ and in the scattering plane as $\left| k, H\right>$ (initial state) or $\left| k, H'\right>$ (scattered state), i.e., we now use $H$ instead of $\hat{\varepsilon}_H$, etc. Our starting point is a maximally entangled state
\begin{equation}\label{BellState}  
\left| i \right> = \frac{1}{\sqrt{2}} \left | 0, a\right>\left( \left| k, H\right>_S \left| q, V\right>_I - \left| k, V\right>_S \left| q, H\right>_I \right),
\end{equation}
the subscript $I$ and $S$ denote the idler and signal photon, respectively, $\left| 0, a \right>$ is the state of the electron at rest with spin $a$ and $q$ is the idler photon momentum. The final state is written as
\begin{widetext} 
\begin{equation}  
\begin{aligned}
\left| f \right> &= \frac{1}{\sqrt{2 N_H}} \big[ \left| 0, a \right>\left| k, H \right>_S \\ 
&+ i (2\pi)^4\sum_{a', \hat{\varepsilon}' }\int_{\vec{p'}, \vec{k'}}\delta^4(k-p'-k')\times M(0, a, k, H; p', a', k',  \hat{\varepsilon}' ) \left| p', a'\right>\left| k',  \hat{\varepsilon}' \right>_S \big] \left| q, V\right>_{I} \\
&- \frac{1}{\sqrt{2 N_V}} \big[ \left| 0, a \right>\left| k, V \right>_S \\
& + i (2\pi)^4\sum_{a',  \hat{\varepsilon}' }\int_{\vec{p'}, \vec{k'}}\delta^4(k-p'-k') \times M(0, a, k, V; p', a', k',  \hat{\varepsilon}' ) \left| p', a'\right>\left| k', \hat{\varepsilon}' \right>_S \big] \left| q, H\right>_{I}. \\
\end{aligned}
\end{equation}
\end{widetext} 
where $\left| p', a'\right>$ is the final state of the scattered electron and $N_H$ and $N_V$ are the normalization constants of the final states if the initial signal photon states are $\left| k, H\right>_S$  and $\left| k, V \right>_S$, respectively. 
\begin{equation}
\begin{aligned}
&N^{'}_H=\frac{1}{8m\omega^2 \mathcal{V}^3}N_H, \text{with} \\
&N_H = 1+\frac{\pi \mathcal{T} \alpha^2}{2m^2 \mathcal{V} } (I_0+I_1+I_2) , \\
&N^{'}_V=\frac{1}{8m\omega^2 \mathcal{V}^3}N_V, \text{with} \\
&N_V = 1+\frac{\pi \mathcal{T} \alpha^2}{2m^2 \mathcal{V}} (J_0+J_1+J_2).
\end{aligned}
\end{equation}
$I_0, I_1, I_2, J_0, J_1,$ and $J_2$ are defined below. The factor $1/8m\omega^2 \mathcal{V}^3$ comes from the anticipation of calculating the reduced density matrix. This way, the reduced density matrix is dimensionless. The final state density matrix is $\rho_{f}=\left| f \right> \left< f \right|$. In order to find the reduced density matrix of photons, we traced out the electron states with 
\begin{equation}
Tr_e[ \rho_f ] = \sum_{b}\int \frac{d^3 p}{(2\pi)^3}\frac{1}{2E_{p}}\left< p, b| \rho_f| p,b\right>
\end{equation}
where $\left| p, b\right>$ is the state of the electron with momentum $p$ and spin $b$. After this step, the electron’s degree of freedom, or the environment’s effect, is integrated out, and what is left is the dynamics of the principal part, the system:
\begin{widetext} 
\begin{equation}\label{Tref}  
\begin{aligned}
Tr_e[ \rho_f ]= 
 \frac{2m\mathcal{V}}{N'_{H}} & \big[ \left| k, H \right>_S \left| q, V \right>_I \prescript{}{I}{\left< q, V\right| } \prescript{}{S}{ \left< k, H \right |} 
+ \frac{1}{2}\sum_{a, b, \hat{\varepsilon}', \hat{\varepsilon}''} \int\frac{d^3 k'}{(2\pi)^3 2\omega'}\frac{2\pi \mathcal{T} \delta(E_i-E_f)}{2E_p} \\
& M(0,a,k,H;p,b,k',\hat{\varepsilon}')M^{*}(0,a,k,H;p,b,k'',\hat{\varepsilon}'')  \left| k', \hat{\varepsilon}' \right>_S \left| q, V \right>_I \prescript{}{I}{\left< q, V\right| } \prescript{}{S}{ \left< k'', \hat{\varepsilon}'' \right |} \big] \\
+\frac{2m\mathcal{V}}{N'_{V}} &\big[ \left| k, V \right>_S \left| q, H \right>_I \prescript{}{I}{\left< q, H\right| } \prescript{}{S}{ \left< k, V \right |} 
+ \frac{1}{2}\sum_{a, b, \hat{\varepsilon}', \hat{\varepsilon}''} \int\frac{d^3 k'}{(2\pi)^3 2\omega'}\frac{2\pi \mathcal{T} \delta(E_i-E_f)}{2E_p} \\
& M(0,a,k,V;p,b,k',\hat{\varepsilon}')M^{*}(0,a,k,V;p,b,k'',\hat{\varepsilon}'')  \left| k', \hat{\varepsilon}' \right>_S \left| q, H \right>_I \prescript{}{I}{\left< q, H\right| } \prescript{}{S}{ \left< k'', \hat{\epsilon}'' \right |} \big]\\
-\frac{2m\mathcal{V}}{\sqrt{N'_{V}N'_{H}}} &\big[ \left| k, V \right>_S \left| q, H \right>_I \prescript{}{I}{\left< q, V\right| } \prescript{}{S}{ \left< k, H \right |} 
+ \frac{1}{2}\sum_{a, b, \hat{\varepsilon}', \hat{\varepsilon}''} \int\frac{d^3 k'}{(2\pi)^3 2\omega'}\frac{2\pi \mathcal{T} \delta(E_i-E_f)}{2E_p} \\
& M(0,a,k,V,b,k',\hat{\varepsilon}')M^{*}(0,a,k,H;p,b,k'',\hat{\varepsilon}'')  \left| k', \hat{\varepsilon}' \right>_S \left| q, H \right>_I \prescript{}{I}{\left< q, V \right| } \prescript{}{S}{ \left< k'', \hat{\varepsilon}'' \right |} \big]\\
-\frac{2m\mathcal{V}}{\sqrt{N'_{V}N'_{H}}} &\big[ \left| k, H \right>_S \left| q, V \right>_I \prescript{}{I}{\left< q, H \right| } \prescript{}{S}{ \left< k, V \right |} 
+ \frac{1}{2}\sum_{a, b, \hat{\varepsilon}', \hat{\varepsilon}''} \int\frac{d^3 k'}{(2\pi)^3 2\omega'}\frac{2\pi \mathcal{T} \delta(E_i-E_f)}{2E_p} \\
& M(0,a,k,H,b,k',\hat{\varepsilon}')M^{*}(0,a,k,V;p,b,k'',\hat{\varepsilon}'')  \left| k', \hat{\varepsilon}' \right>_S \left| q, V \right>_I \prescript{}{I}{\left< q, H \right| } \prescript{}{S}{ \left< k'', \hat{\varepsilon}'' \right |} \big] .
\end{aligned}
\end{equation}
To make the expression more compact, we further trace out the momenta of the signal photon and the idler photon with $\int\frac{d^3 k'}{(2\pi)^3}\frac{1}{2\omega'} \left<k' | \rho_f | k'\right>$. The result is
\begin{equation}\label{Trrhof} 
\begin{aligned}
Tr_e[ \rho_f ]&=  \\
 \frac{1}{2N_{H}} &\left[ \left| H \right>_S \left| V \right>_I \prescript{}{I}{\left< V\right| } \prescript{}{S}{ \left< H \right |} 
+ \frac{\mathcal{T}}{16m^2 \mathcal{V}} \int\frac{d \Omega'}{4\pi^2}\left(\frac{\omega' }{\omega}\right)^2 \sum_{\hat{\varepsilon}', \hat{\varepsilon}''}  M(H\rightarrow \hat{\varepsilon}')M^{*}(H\rightarrow \hat{\varepsilon}'') \left| \hat{\varepsilon}' \right>_S \left| V \right>_I \prescript{}{I}{\left< V\right| } \prescript{}{S}{ \left< \hat{\varepsilon}'' \right |} \right] \\ 
+\frac{1}{2N_{V}} &\left[  \left| V \right>_S \left| H \right>_I \prescript{}{I}{\left< H\right| } \prescript{}{S}{ \left< V\right |} 
+ \frac{\mathcal{T}}{16m^2 \mathcal{V}} \int\frac{d \Omega'}{4\pi^2}\left(\frac{\omega' }{\omega}\right)^2 \sum_{\hat{\varepsilon}', \hat{\varepsilon}''}  M(V\rightarrow \hat{\varepsilon}')M^{*}(V\rightarrow \hat{\varepsilon}'') \left| \hat{\varepsilon}' \right>_S \left| H \right>_I \prescript{}{I}{\left< H \right| } \prescript{}{S}{ \left< \hat{\epsilon}'' \right |} \right]\\ 
-\frac{1}{2\sqrt{N_{V}N_{H}}} & \left[ \left| V \right>_S \left| H \right>_I \prescript{}{I}{\left< V\right| } \prescript{}{S}{ \left< H \right |} 
+ \frac{\mathcal{T}}{16m^2 \mathcal{V}} \int\frac{d \Omega'}{4\pi^2}\left(\frac{\omega' }{\omega}\right)^2 \sum_{\hat{\varepsilon}', \hat{\varepsilon}''}  M(V\rightarrow \hat{\varepsilon}')M^{*}(H\rightarrow \hat{\varepsilon}'') \left| \hat{\varepsilon}' \right>_S \left| H \right>_I \prescript{}{I}{\left< V\right| } \prescript{}{S}{ \left< \hat{\varepsilon}'' \right |} \right]\\ 
-\frac{1}{2\sqrt{N_{V}N_{H}}} &\left[  \left| H \right>_S \left| V \right>_I \prescript{}{I}{\left< H\right| } \prescript{}{S}{ \left< V \right |} 
+ \frac{\mathcal{T}}{16m^2 \mathcal{V}} \int\frac{d \Omega'}{4\pi^2}\left(\frac{\omega' }{\omega}\right)^2 \sum_{\hat{\varepsilon}', \hat{\varepsilon}''}  M(H\rightarrow \hat{\varepsilon}')M^{*}(V\rightarrow \hat{\varepsilon}'') \left| \hat{\varepsilon}' \right>_S \left| V \right>_I \prescript{}{I}{\left< H \right| } \prescript{}{S}{ \left< \hat{\varepsilon}'' \right |} \right]  \\ 
\end{aligned}
\end{equation}
where $M(\varepsilon\rightarrow \varepsilon')$ denotes the $M$ matrix of the transition from the photon with polarization $\varepsilon$ to the photon with polarization $\varepsilon'$ that has only two types, $H$ and $V$. The simplification procedure for $Tr_e[\rho_f]$ is straightforward but very lengthy. The details are in Appendix B \cite{{CHLSM}}. We present the resulting reduced density matrix with rows and columns ordered as $\left | H \right>_S \left | H \right>_I$, $\left | H \right>_S \left | V \right>_I$, $\left | H \right>_S \left | Z \right>_I$, $\left |V \right>_S \left | H \right>_I$,  $\left | V \right>_S \left | V \right>_I$, $\left | V \right>_S \left | Z \right>_I$,  $\left | Z \right>_S \left | H \right>_I$, $\left | Z \right>_S \left | V \right>_I$, and $\left | Z \right>_S \left | Z \right>_I$, where $\left | Z \right>$ denotes the polarization along the $z$-axis direction.
\begin{equation}\label{DM99}  
Tr_e[ \rho_f ]=\frac{1}{2}\left(
\begin{array}{ccc ccc ccc}
 \frac{ \mathcal{T} \chi }{ \mathcal{V} N_V} J_1&0 & 0  &0 & \frac{- \mathcal{T} \chi }{ \mathcal{V} \sqrt{N_H N_V}} J_1 & 0 & \frac{  \mathcal{T}\chi }{ \mathcal{V} N_V}J_3 &0 & 0 \\ 
0 & \frac{1+ \mathcal{T} \chi I_1/ \mathcal{V}}{N_H}  & 0  & - \frac{1 + \mathcal{T} \chi L_1/ \mathcal{V}}{ \sqrt{N_H N_V}} & 0 & 0 &0 & \frac{ \mathcal{T} \chi }{ \mathcal{V} N_H} I_3 & 0  \\
0 & 0 & 0  & 0 & 0 & 0 &0 &0 & 0  \\
0 & -  \frac{1 + \mathcal{T} \chi L_1/ \mathcal{V}}{ \sqrt{N_H N_V}}  & 0  & \frac{1+ \mathcal{T} \chi J_0/ \mathcal{V}}{N_V}& 0 & 0 &0 & \frac{ -\mathcal{T} \chi }{ \mathcal{V} \sqrt{N_H N_V}} L_2& 0  \\
 \frac{-T \chi }{V \sqrt{N_H N_V}} J_1&0 & 0  &0 &  \frac{ \mathcal{T} \chi }{ \mathcal{V} N_H} I_0 & 0 &\frac{- \mathcal{T} \chi }{ \mathcal{V} \sqrt{N_H N_V}} J_3 & 0 & 0 \\ 
0 & 0 & 0  & 0 & 0 & 0 &0 &0 & 0  \\
\frac{ \mathcal{T}\chi}{ \mathcal{V} N_V} J_3 & 0 & 0  & 0 & \frac{ -\mathcal{T} \chi }{ \mathcal{V}\sqrt{N_H N_V}} J_3 & 0 &  \frac{ \mathcal{T} \chi }{ \mathcal{V}N_V} J_2 &0 & 0  \\
0 & \frac{ \mathcal{T}\chi}{ \mathcal{V} N_H}  I_3 & 0  & \frac{ -\mathcal{T} \chi}{ \mathcal{V} \sqrt{N_H N_V}} L_2  & 0 & 0 &0 &\frac{ \mathcal{T} \chi }{ \mathcal{V} N_H} I_2& 0  \\
0 & 0 & 0  & 0 & 0 & 0 &0 &0 & 0  \\
\end{array}
\right),
\end{equation}
where $\chi=\frac{\pi\alpha^2}{2m^2} = \frac{3\sigma_t}{16}$. The integrals $I's$, $J's$, and $L's$ are related to the Klein-Nishina’s cross section formula. We define 
\begin{equation}\label{Fth}
F(\theta, \theta')=\frac{4}{r^2_0}\frac{d\sigma}{d\Omega}=\left(\frac{1}{1+\beta(1-\cos \theta)} \right)^{2} \left(  \frac{1}{1+\beta(1-\cos \theta)} + (1+\beta(1-\cos \theta) -2 + 4 \cos^{2}\theta' \right),
\end{equation}
where $\beta\equiv \frac{\omega}{m}$, the argument $\theta$ is equal to the angle between the wave vectors of the incident and scattered photons, and $\theta'$ is equal to the angle between the polarizations of the incident and scattered photons. \\
 \end{widetext}
\begin{equation}\label{I0}
I_{0}=\int^{\pi}_{0} F(\theta, \frac{\pi}{2}) \sin \theta d\theta,
\end{equation}
\begin{equation}\label{I1}
I_{1}=\int^{\pi}_{0} F(\theta, \theta'=\theta) \cos^2 \theta \sin \theta d\theta ,
\end{equation}
\begin{equation}\label{I2}
I_{2}=\int^{\pi}_{0} F(\theta, \theta'=\theta)  \sin^3 \theta d\theta ,
\end{equation}
\begin{equation}\label{I3}
I_{3}=\int^{\pi}_{0} F(\theta, \theta'=\theta) \cos\theta \sin^2 \theta d\theta ,
\end{equation}
\begin{equation}\label{J0}
J_{0}=\int^{\pi}_{0} F(\theta, 0) \sin \theta d\theta ,
\end{equation}
\begin{equation}\label{J1}
J_{1}=\int^{\pi}_{0} F(\theta, \frac{\pi}{2}) \cos^2 \theta \sin \theta d\theta ,
\end{equation}
\begin{equation}\label{J2}
J_{2}=\int^{\pi}_{0} F(\theta, \frac{\pi}{2})  \sin^3 \theta d\theta ,
\end{equation}
\begin{equation}\label{J3}
J_{3}=\int^{\pi}_{0} F(\theta, \frac{\pi}{2}) \cos\theta \sin^2 \theta d\theta ,
\end{equation}
\begin{equation}\label{L1}
L_{1}=\int^{\pi}_{0} F(\theta, 0) \cos^2 \theta \sin \theta d\theta ,
\end{equation}
\begin{equation}\label{L2}
L_{2}=\int^{\pi}_{0} F(\theta, 0)  \cos\theta \sin^2 \theta d\theta ,
\end{equation}
The total cross section of the Compton scattering $\sigma_t=\frac{\chi}{2}(I_0+I_1+I_2+J_0+J_1+J_2)$. The integrals from Eqs. (\ref{I0})-(\ref{L2}) can be evaluated exactly; see Appendix D \cite{{CHLSM}} for details. \\
\begin{figure}
    \centering
    \subfigure[]{\includegraphics[width=0.37\textwidth]{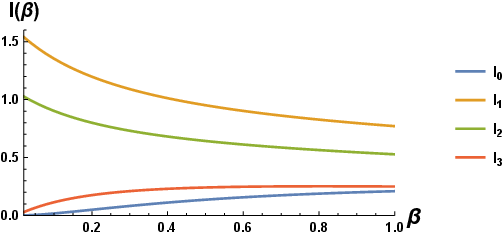}} 
    \subfigure[]{\includegraphics[width=0.37\textwidth]{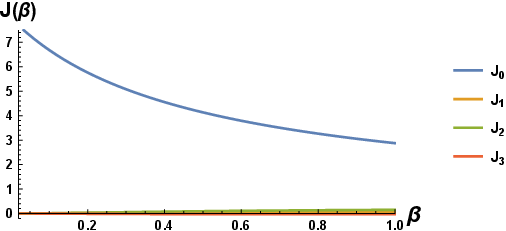}} 
    \subfigure[]{\includegraphics[width=0.37\textwidth]{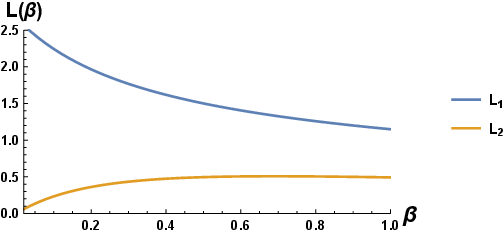}} 
    \caption{The results of integration of (a) $I_0\sim I_3$ (b) $J_0\sim J_3$ (c) $L_1, L_2$ with respect to different $\beta$. The relation $J_0=3 (I_0+I_1)$ in the Thomson limit is satisfied. }
    \label{figIntegrals}
\end{figure}
The dimensionless $I's, J's$, and $L's$  are computed, and the results are shown in Fig. \ref{figIntegrals} as functions of $\beta$. It is noticed that $J_0$ has the largest magnitude, followed by $L_1$, and then $I_1 \sim I_3$. $J_0$ is the integral of $| M^{*}(V \rightarrow V)|^2$, $I_1, I_2$ and $I_3$ are related to $| M(H \rightarrow H')|^2$, and $L_1$ is that of $M(H \rightarrow H')M^{*}(V \rightarrow V)$. Clearly, $J_0$ has the largest magnitude because of the alignment of the polarizations before and after scattering. $J_0, I_1$, and $I_2$ directly contribute to the diagonal matrix elements. In other words, they are related to the probabilities of the final states, while $L_1$ is related to the correlation between $\left| H' \right>$ and $\left| V \right>$. \\
\indent
The final density matrix in Eq. (\ref{DM99}) is very interesting. First, its third, sixth, and ninth rows and columns are zero. This is because the idler photon's polarization does not have a $z$-component. Secondly, the density matrix is block-diagonalized, with rows and columns 2, 4, and 8 grouped together and 1, 5, and 7 grouped together. The former group forms a $3\times 3$ matrix in which the polarization of the signal photon remains unchanged after scattering ($V$ to $V$ and $H$ to $H’$). The latter group forms another $3\times 3$ matrix in which the polarization of the signal photon is flipped ($V$ to $H’$ and $H$ to $V$). The matrix elements of the latter are smaller in magnitude than the scattered part of the former. It is a consequence of Compton scattering, which favors the initial and final polarizations being aligned (see Eq. (\ref{KNformula})). \\
\indent 
\begin{widetext}
One thing stands out is the effect of Compton scattering on entanglement. The state $ ( \left| H \right>_S \left| H \right>_I + \left| V \right>_S \left| V \right>_I )/\sqrt{2} $ has the largest probability of occurrence compared to two other Bell states if the initial state is $ ( \left| H \right>_S \left| V \right>_I - \left| V \right>_S \left| H \right>_I )/\sqrt{2} $. To see this, we focus on the matrix formed by rows and columns labeled 2, 4, and 8. The corresponding initial density matrix is
\begin{equation}
\rho_{i, 248}=\frac{1}{2}\left(
\begin{array}{ccc}
1 & -1 & 0 \\
-1 & 1& 0 \\
0  & 0 & 0\\
\end{array}
\right).
\end{equation}
It can be diagonalized by
\begin{equation}
U=\left(
\begin{array}{ccc}
\frac{1}{\sqrt{2}} & -\frac{1}{\sqrt{2}} & 0 \\
-\frac{1}{\sqrt{2}} & \frac{1}{\sqrt{2}}& 0 \\
0  & 0 & 1 \\
\end{array}
\right) 
\end{equation}
as
\begin{equation}
 \rho'_{i, 248} = U \rho_{i, 248} U^{-1} = 
 \left(
\begin{array}{ccc}
1 & 0 & 0 \\
0 & 0 & 0 \\
0  & 0 & 0 \\
\end{array}
\right). 
 \end{equation} 
 The order of rows of the new matrix is $HV-VH$, $HV+VH$, and $ZV$. It indicates that our initial state of photons is $ ( \left| H \right>_S \left| V \right>_I - \left| V \right>_S \left| H \right>_I )/\sqrt{2} $. We apply the same transformation to the 2, 4, and 8 rows and columns of the final density matrix and obtain
 \begin{equation}
 \begin{aligned}
 \rho'_{i, 248}& = U \rho_{i, 248} U^{-1} \approx \\
& \left(
\begin{array}{ccc}
1- \frac{\mathcal{T} \chi}{ \mathcal{V}} \left( \frac{3I_0+I_1+3I_2+J_0+3J_1+3J_2-2L_1}{2}  \right) & \frac{1 + \mathcal{T} \chi I_1/\mathcal{V} }{N_H} -\frac{1 + \mathcal{T} \chi J_0/\mathcal{V} }{N_V}  & \frac{\sqrt{2} \mathcal{T} \chi }{\mathcal{V} } (I_3+L_2) \\
 \frac{1 + \mathcal{T} \chi I_1/\mathcal{V} }{N_H} -\frac{1 + \mathcal{T} \chi J_0/\mathcal{V} }{N_V}  & \frac{\mathcal{T} \chi}{ \mathcal{V}} \left( \frac{J_0 - J_1- J_2- I_0+I_1-I_2-2L_1}{2}  \right) &  \frac{\sqrt{2} \mathcal{T} \chi }{\mathcal{V} } (I_3-L_2)  \\
 \frac{\sqrt{2} \mathcal{T} \chi }{\mathcal{V} } (I_3+L_2)   &  \frac{\sqrt{2} \mathcal{T} \chi }{\mathcal{V} } (I_3-L_2)  & 0 \\
\end{array}
\right),
 \end{aligned}
 \end{equation} 
 \end{widetext}
where we kept only the zeroth- and first-order terms in $\chi$. Clearly, the state  $\Psi^{-} \equiv  ( \left| H \right>_S \left| V \right>_I -  \left| V \right>_S \left| H \right>_I )/\sqrt{2} $  has the largest diagonal element due to the unscattered part. However, as shown in FIG. \ref{figIntegrals}, the diagonal element $( \rho'_{f ,248} )_{22}$ is an order of magnitude larger than those of the diagonal elements $\rho_{11}$ (denoting $HH$), $\rho_{55}$ (denoting $VV$), and $\rho_{77}$ (denoting $ZH$), in Eq. (\ref{DM99}). This indicates that the probability of  $\Psi^{+} \equiv  ( \left| H \right>_S \left| V \right>_I +  \left| V \right>_S \left| H \right>_I )/\sqrt{2} $ emerging from scattering is the largest compared to states $ \Phi^{+} \equiv  ( \left| H \right>_S \left| H \right>_I +  \left| V \right>_S \left| V \right>_I )/\sqrt{2} $ and $ \Phi^{-} \equiv  ( \left| H \right>_S \left| H \right>_I -  \left| V \right>_S \left| V \right>_I )/\sqrt{2} $. It can be inferred that Compton scattering also favors transitions between $ \Phi^{+} $ and $ \Phi^{-} $ but not inter-pair transitions. \\
\indent
The physical reason why the Compton scattering favors transitions between $ \Psi^{-} $ and $ \Psi^{+} $ is that the signal photon with polarization $V$ will mostly be scattered to $V$ and $H$ mostly to $H’$, and the entangled polarization states will mostly transition to each other in the same Bell-state subgroup. This character is also demonstrated in the Kraus operator formulation in section IV. Consequently, the effect of Compton scattering is approximately a Pauli matrix $-\sigma_z$ multiplied by $\alpha^2$ acting on the signal photon state.
\section{The Kraus operators, entropy and mutual information}
Quantum channels effectively describe the net effects of the interaction between the principal system and its environment on the principal system. Exploring the physics of the interaction between the principal system and its environment is clearer in the quantum-channel representation. After developing the fundamental theory of Compton scattering in the single-photon and entangled-photon cases, we present our calculations in Sec. II and III in the Kraus operator format.\\
\indent
We shall consider only the large electron mass limit for which the visible light and most EM waves of application are excellent approximations. In this limit, $\omega'\approx\omega$ and the Klein-Nishina differential cross section reduces to that of Thomson scattering. One quickly notices that the cross section vanishes if the $\hat{\varepsilon}$ and $\hat{\varepsilon}'$ are orthogonal. Therefore, it is advantageous to choose $\hat{\varepsilon}_V$ and $\hat{\varepsilon}_H$ as basis where $\hat{\varepsilon}_V$ and $\hat{\varepsilon}_H$ are perpendicular to each other. Furthermore, the incident photon with polarization $\hat{\varepsilon}_V$ can only be scattered into a photon with the same polarization. The incident photon with polarization  $\hat{\varepsilon}_H$ can only be scattered into a photon with polarization   $\hat{\varepsilon}_{H'}$, which is perpendicular to the scattered photon momentum. This is the physical origin of the depolarization of the horizontal polarization.\\
\indent
As mentioned, we focus only on the photons’ polarization degree of freedom. We first consider single-photon Compton scattering. The Kraus operators associated with Compton scattering can be calculated using a superposition state as the initial and final states, as derived in Sec. II. We use a superposition state as the initial state to comprehensively cover all the basis states, correlations, and quantum channels. \\
\indent
We use this state because it probes the full interaction channels. We use a maximally entangled state as the initial state in Compton scattering of entangled photons. In principle, the Choi matrix method can systematically derive the Kraus operators \cite{Choi}. A simple algebraic manipulation suffices to derive the Kraus operators in this case. We define $p = \frac{32}{15} \frac{\chi \mathcal{T}}{\mathcal{V}}$ for the simplicity. Note that $p$ is a dimensionless, small positive number of order $\alpha^2$, the coefficient of the first order of QED perturbation theory. In view of Eq. (\ref{initialSinglePhoton}) and Eq. (\ref{DMSinglePhoton}), they give the density matrix of the initial state and the final state, which reads
\begin{equation} 
\begin{aligned}
&\rho^{sup}_{i}=\frac{1}{2} \left(
\begin{array}{ccc }
1 & 1 & 0    \\
1 & 1 & 0    \\
0 & 0 & 0   \\
\end{array}
\right),\\
&\rho^{sup}_{f}=\frac{1}{2} \left(
\begin{array}{ccc }
1-\frac{p}{2} & 1-\frac{5p}{4} & 0    \\
1-\frac{5p}{4}  & 1 & 0    \\
0 & 0 & \frac{p}{2}  \\
\end{array}
\right).
\end{aligned}
\end{equation}
The Kraus operators, see Appendix E of Ref. \cite{CHLSM}, read
\begin{equation}  
\begin{aligned}
&K_0= \left(
\begin{array}{ccc }
\sqrt{1-\frac{p}{2}} & 0 & 0    \\
0 & \frac{1-5p/4}{{\sqrt{1-p/2}}} & 0    \\
0 & 0 & \sqrt{1-\frac{p}{2}}   \\
\end{array}
\right), \\
&K_1=\left(
\begin{array}{ccc }
0 & 0 & \sqrt{\frac{p}{2}}   \\
0 & 0 & 0    \\
 \sqrt{\frac{p}{2}}  & 0 & 0  \\
\end{array}
\right),\\
&K_2=\left(
\begin{array}{ccc }
0 & 0 &0  \\
0 &   \sqrt{\frac{2p}{1-p/2}} & 0    \\
0  & 0 & 0  \\
\end{array}
\right).
\end{aligned}
\end{equation}
If they are expressed in Gell-Mann matrix basis, $K_0$ and $K_2$ are the linear combination of $\lambda_3$, $\lambda_8$ and identity matrix $\mathcal{I}$; $K_1$ is proportional to $\lambda_4$. The combined action of $K_0$ and $K_2$ depolarizes $H$ but leaves $V$ intact; it is also responsible for normalizing the correlation terms; $K_1$ partially tilts the $x$-component of $H$ toward the $z$-direction. Note again that the Kraus operators derived here are valid only in the Thomson limit, which is sufficient to demonstrate the physical characteristics. For other energy scopes, for example, in astrophysics (discussed in the next section), the Kraus operators can be obtained in the same way we demonstrated in the Thomson limit case.\\
\indent
The von Neumann entropy of the initial state can be derived from eigenvalues $\mu_j$, the eigenvalues of the density matrix. Initially, 
\begin{equation}\label{entropyi}  
S_i=-\sum_{j}\mu_{j} \ln \mu_{j} =0.
\end{equation} 
As the initial state is a pure state. The entropy of the final state is
\begin{equation}\label{entropyf}  
\begin{aligned}
S_{f}&= -\sum_{j}\mu_{j} \ln \mu_{j} \\
& = -\frac{p}{4}\ln \frac{p}{4} - \left(1-\frac{3p}{4}\right) \ln\left(1-\frac{3p}{4}\right) - \frac{p}{2}\ln \frac{p}{2} \\
&  \approx - \frac{3p}{4}\ln p,
\end{aligned}
\end{equation}
where we assumed $-p\ln p \gg p$. The entropy increases because of the depolarization of the $H$ state.\\
\indent
For the Compton scattering of an entangled pair of photons, we choose the Bell state as the initial state, as in Eq. (\ref{BellState}). It suffices to cover all basis states, correlations, and quantum channels. The initial and the final density matrix read
\begin{equation}\label{initialDM}
\rho_{i}=\frac{1}{2} \left(
\begin{array}{ccc ccc ccc}
0 &0 & 0  &0 & 0& 0 & 0 &0 & 0 \\ 
0 & 1 & 0  & -1 & 0 & 0 &0 &0 & 0  \\
0 & 0 & 0  & 0 & 0 & 0 &0 &0 & 0  \\
0 & -1 & 0  & 1 & 0 & 0 &0 &0 & 0  \\
0 &0 & 0  &0 & 0 & 0 &0 &0 & 0 \\ 
0 & 0 & 0  & 0 & 0 & 0 &0 &0 & 0  \\
0 & 0 & 0  & 0 & 0 & 0 &  0 &0 & 0  \\
0 & 0 & 0  & 0 & 0 & 0 &0 &0 & 0  \\
0 & 0 & 0  & 0 & 0 & 0 &0 &0 & 0  \\
\end{array}
\right),
\end{equation}
and in view of Eq. (\ref{DM99})
\begin{equation}\label{finalDM}
\rho_{f}=\frac{1}{2} \left(
\begin{array}{ccc ccc ccc}
0 &0 & 0  &0 & 0& 0 & 0 &0 & 0 \\ 
0 & 1-\frac{p}{2} & 0  & -1+\frac{5p}{4}  & 0 & 0 &0 &0 & 0  \\
0 & 0 & 0  & 0 & 0 & 0 &0 &0 & 0  \\
0 & -1+\frac{5p}{4} & 0  & 1 & 0 & 0 &0 &0 & 0  \\
0 &0 & 0  &0 & 0 & 0 &0 &0 & 0 \\ 
0 & 0 & 0  & 0 & 0 & 0 &0 &0 & 0  \\
0 & 0 & 0  & 0 & 0 & 0 &0 &0 & 0  \\
0 & 0 & 0  & 0 & 0 & 0 &0 &  \frac{p}{2}& 0  \\
0 & 0 & 0  & 0 & 0 & 0 &0 &0 & 0  \\
\end{array}
\right).
\end{equation}
\indent
It is not difficult to observe that the structures of the initial and final density matrices of the current case are similar to those of the single-photon Compton scattering case. The Kraus operators, in the current case, can be expressed by $K_{i} \otimes\mathcal{I}$, where $K_{i}$'s are the Kraus operators derived for the single-photon Compton scattering case; they act on the polarization space of the signal photon, while $\mathcal{I}$ is the identity matrix that acts on the polarization space of the idler photon. Since the initial state we choose is a maximally entangled state, the Kraus operators we derived fully represent all the quantum channels related to Compton scattering and can be applied to all the initial states \cite{Nielsen, Choi, Jamiolkowski}.\\
\indent
The initial and final reduced density matrices of the signal photon and the idler photon are given by
\begin{equation}  
\begin{aligned}
&\rho_{S,i}= \frac{1}{2}\left(
\begin{array}{ccc }
1 & 0 & 0    \\
0 & 1 & 0    \\
0 & 0 & 0 \\
\end{array}
\right),
\quad
\rho_{S,f}= \frac{1}{2}\left(
\begin{array}{ccc }
1-\frac{p}{2} & 0 & 0    \\
0 & 1 & 0    \\
0 & 0 & \frac{p}{2} \\
\end{array}
\right), \\
&\rho_{I,i}=\frac{1}{2} \left(
\begin{array}{ccc }
1 & 0 & 0   \\
0 & 1 & 0    \\
0  & 0 & 0  \\
\end{array}
\right),
\quad
\rho_{I,f}=\frac{1}{2} \left(
\begin{array}{ccc }
1 & 0 & 0   \\
0 & 1 & 0    \\
0  & 0 & 0  \\
\end{array}
\right).
\end{aligned}
\end{equation} 
The initial state is a maximally entangled state and hence has zero entropy. The final entanglement entropy is the same as that of the single-photon Compton scattering case. The increase of entanglement entropy is also due to the depolarization of the $H$ state.\\
\indent
The mutual information of the system consisting of an entangled signal photon and an idler photon is defined as
\begin{equation}
I(S:I)=S(\rho_S)+S(\rho_I)-S(\rho_{SI}),
\end{equation}
where $I(S:I)$ is the mutual information, $S (\rho_S)$ is the entropy of the reduced density matrix of the signal photon, and $S(\rho_I)$ is the entropy of the idler photon, $S(\rho_{SI})$ is the entropy of the system. The initial mutual information is  $I_{i}(S:I)=2\ln 2$. The final mutual information is
\begin{equation} 
\begin{aligned}
I_{f}(S:I) &=S_{f}(\rho_S)+S_{f}(\rho_I)-S_{f}(\rho_{SI}) \\
& = 2 \ln 2 - \frac{p}{4}\ln p + \frac{3p}{4} \ln p \\
& = 2 \ln 2 + \frac{p}{2}\ln p  \\
& < I_i (S:I)
\end{aligned}
\end{equation}
The final mutual information decreases slightly by $\frac{p}{2}\ln p$ because the Compton scattering weakens the entanglement correlation. From the form of the information loss, it is evident that the reduction in mutual information primarily arises from the depolarization of the $H$ state. It was initially confined to the $x$-axis. After Compton scattering, it disperses into the $xz$ plane, becoming the primary source of information loss. \\
\indent
It is noted that the idler photon's reduced density matrix remains unchanged after the Compton scattering. This is consistent with the notion that QED is a local interaction and should not affect the idler photon, which does not participate in the interaction. Our result shows that the reduced density matrix of the idler photon remains invariant after Compton scattering. This is compatible with Shivashankara's calculation of unitarity processes \cite{Shanmuka}. Attempts to use the reduced density matrix of the idler photon alone to probe the interaction acting on the signal photon \cite{Araujo} contradict physical intuition. However, as indicated in the final density matrix of the entangled photons, the measurement of the correlation terms indeed changes. This supports the assertion that the correlation or superposition of entangled photons can enhance quantum detections \cite{Torrome}.\\
\section{Discussions and Applications}
\indent
We can compare our results with those of Caradonna \cite{Caradonna2}, expressed in terms of Stokes parameters. Our initial state in Eq. (\ref{initialDM}) gives a density matrix 
\begin{equation} 
\rho_i=\frac{1}{2} \left(
\begin{array}{cccc}
0 & 0 & 0  &0\\ 
0 & 1 & -1  & 0  \\
0 & -1 & 1  & 0   \\
0 & 0 & 0  & 0   \\
\end{array}
\right).
\end{equation}
in the $\left| R \right>$ (right-handed) and $\left| L \right>$ (left-handed) representation The corresponding Stokes density matrix is 
\begin{equation} 
S(\rho_i)=\frac{1}{2} \left(
\begin{array}{cccc}
1 & 0 & 0  &0\\ 
0 & -1 & 0  & 0  \\
0 & 0 & 1  & 0   \\
0 & 0 & 0  & -1  \\
\end{array}
\right).
\end{equation}
which results in 
\begin{equation}
S_S= \left(
\begin{array}{c}
1 \\ 
0   \\
0    \\
0  \\
\end{array}
\right), 
\quad
S_I= \left(
\begin{array}{c}
1 \\ 
0   \\
0    \\
0  \\
\end{array}
\right).
\end{equation}
where the subscripts $S$ and $I$ denote the signal and idler photons. The Stokes density matrix of the final state can be obtained by Eq. (31) of Ref. \cite{Caradonna}:
\begin{equation}
S(\rho_f)=S(\rho_i)M(-\phi)T(\theta),
\end{equation}
where $M(\phi)$ is a matrix of rotating system around $z$-axis by angle by angle $-2\phi$ and $T(\theta)$ is a symmetric 4 by 4 matrix of Compton scattering for Stokes parameters given by Fano \cite{Fano}. Although our resulting density matrix is independent of the azimuthal angle $\phi$, applying our results to data analysis is straightforward. Compared to FIG. 2 of Ref. \cite{Caradonna2}, our $V$ direction coincides with its $x'$ axis and our $H'$ with its $y'$ axis. $\phi$ is the angle at which the scattered photon emerges. The initial coordinates $x$ and $y$ can be set arbitrarily as long as they are perpendicular to the propagation direction of the incident photon. This is due to the independence of the scattering cross section from the asymmetrical angle.
\begin{widetext}
\begin{equation} 
\begin{aligned}
T(\theta) &=
\frac{r^2_0}{2}\left( \frac{k'}{k} \right)^2 \left(
\begin{array}{cccc}
\frac{k'}{k}+ \frac{k}{k'}-\sin^2\theta & -\sin^2\theta & 0  &0\\ 
 -\sin^2\theta & 1+\cos^2\theta & 0  & 0  \\
0 & 0 & 2\cos\theta  & 0   \\
0 & 0 & 0  & \left(\frac{k'}{k}+\frac{k}{k'}\right) \cos\theta  \\ 
\end{array}
\right)\\
&=\frac{r^2_0}{2}\left( \frac{k'}{k} \right)^2 (t_{\alpha \beta})_{\alpha,\beta=0,1,2,3}.
\end{aligned}
\end{equation}
 \end{widetext}
 The Stokes vectors after scattering are
 \begin{equation}
S'_S= t_{00} \left(
\begin{array}{c}
1 \\ 
-p' \cos 2\phi   \\
-p' \sin 2\phi     \\
0  \\
\end{array}
\right), 
\quad
S'_I= t_{00}\left(
\begin{array}{c}
1 \\ 
p'   \\
0    \\
0  \\
\end{array}
\right),
\end{equation}
where $p'=\frac{t_{10}}{t_{00}}$. It shows that the degrees of polarization of the signal and idler photons are the same after scattering. The polarization symmetry between the signal and idler photons is preserved after Compton scattering in agreement with Ref. \cite{Caradonna}.\\
\begin{figure}[t]
\centering
\includegraphics[scale=0.24]{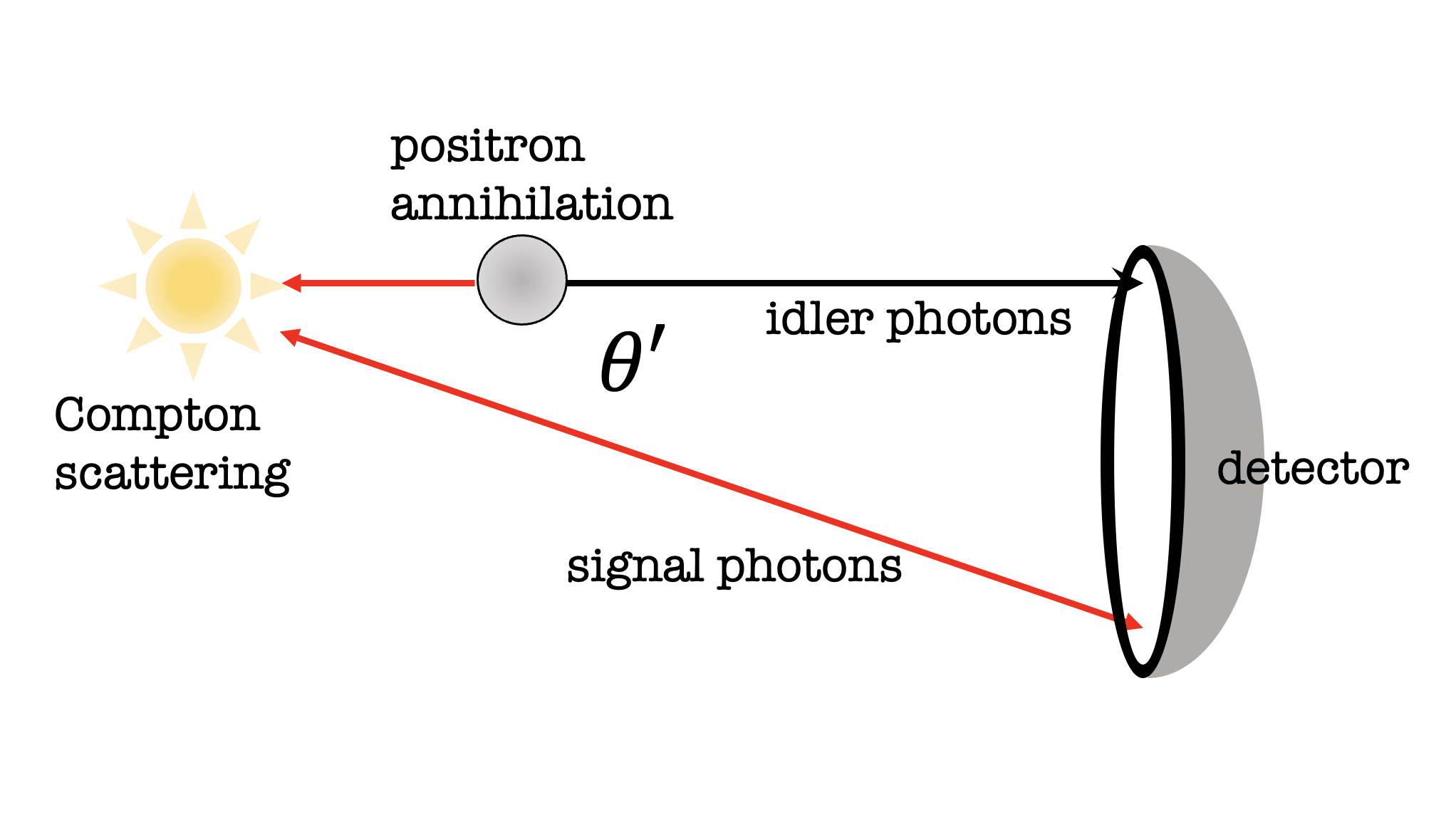}
\caption{ The illustration of the possible signal-idler photon pairs generated by a positron annihilation in astronomy. The idea is the same with the quantum channel application. The detector collects idler and reflected signal photons.}
\label{figastro}
\end{figure}
\indent
Quantum entanglement between two photons can be applied to those produced naturally in astrophysical events, even if they have traveled a long distance. For example, positron annihilations occur in the Sun’s corona or distant galaxies \cite{Deng, Arjun}. The resulting photon pair has a signature $0.51$ MeV spectral line. However, entanglement can only be investigated if both photons are detected. It is difficult because the photons propagate in almost opposite directions. This is where our work can play a crucial role. If one of the photons (the signal photon) is Compton-scattered and changes direction, it is possible for a detector on Earth to receive both photons, as shown in FIG. \ref{figastro}. \\
\indent 
The probability of detecting the signal photon if the detector has received the idler photon is proportional to that of the former being scattered into the angle subtended by the diameter of the detector D: $\theta' = D/L \ll 1$. Here, $\theta'=\pi - \theta \approx 0$, where $\theta$ is the Compton scattering angle, and L is the distance of the annihilation event to Earth. Like the $0.51$ MeV idler photon being the signature of positron annihilation, the $0.17$ MeV signal photon is the signature of backward Compton scattering, in view of Eq. (2). The $0.51$ MeV idler photon and the $0.17$ MeV signal photon can be used to check for entanglement.\\
\section{Conclusion}
Using the classic case of QED, we demonstrate how to derive the net effects of the interaction between an entangled pair of photons and electrons. The normalization of the final density matrix, element by element, is critical for obtaining the correct result. \\
\indent
We derive a complete set of Kraus operators for Compton scattering in the Thomson limit, using the initial and final density matrices obtained from the fundamental theory calculation. The form of the Kraus operators indicates that the Compton scattering depolarizes the $H$ state, tilting the $H$ polarization to the original photon momentum direction while keeping the $V$ polarization unchanged. The increase of entanglement entropy and the decrease of mutual information arise for the same reason.\\
\indent
The reduced density matrix of the idler photon remains intact, as expected. Compton scattering is a local interaction; it should not affect a particle that does not participate in it. Thus, attempting to measure the idler photon alone to enhance quantum sensing efficiency contradicts fundamental principles of physics. \\
\indent
We present our results in terms of the Stokes matrix and show that the degree of polarization of the signal photon and idler photon is the same after scattering in the entire energy range. The study of entanglement produced by astrophysical sources is investigated. The photon pair created by electron-positron annihilation is analyzed. The Compton-scattered signal photon has an energy of $0.17$ MeV, a clear signature of entanglement.
\section{Acknowledgment}
This work is financially supported by the Quantum Information Center, Chung Yuan Christian University.

\end{document}